\documentclass[letterpaper, preprint, paper,11pt]{AAS}	% for preprint proceedings

\usepackage{bm}
\usepackage{amsmath}
\usepackage{subfigure}
\usepackage[colorlinks=true, pdfstartview=FitV, linkcolor=black, citecolor= black, urlcolor= black]{hyperref}
\usepackage{overcite}
\usepackage{footnpag}			      	% make footnote symbols restart on each page
\usepackage{multirow}
\PaperNumber{23-140}

\begin{document}

\title{Interstellar Object Uncertainty Evolution and Effect on Fast Flyby Delivery and Required Delta-V}

\author{Declan M. Mages\thanks{Mission Design and Navigation, Jet Propulsion Laboratory, California Institute of Technology.},  
\ Davide Farnocchia\footnotemark[1],
\ Benjamin Donitz\thanks{Project Systems Engineering and Formulation, Jet Propulsion Laboratory, California Institute of Technology.}
}

\maketitle{} 		

\begin{abstract}
Interstellar objects (ISOs) are small bodies that can travel through our solar system from other star systems. When present in our solar system, they represent an opportunity to study the properties and origins of these objects, as well as the potential for cross-pollination of material between star systems. With current propulsion technology, rendezvous with these objects is likely infeasible, and thus the maximum science return results from a rapid response flyby and impactor. However, while trajectories to ISOs may be feasible, their potentially high ephemeris uncertainties and high-speed hyperbolic orbits present significant challenges to navigation. In this paper we assess these challenges by modeling the uncertainties of reachable synthetic ISOs as a function of time, as derived by measurements from ground observatories and an approaching spacecraft. From these uncertainties we derive the final delivery accuracy of fast flyby spacecraft to the ISO and required statistical delta-v for navigation. We find that these two challenges can lead to hundreds of meters-per-second or even kilometers-per-second of required statistical delta-v for navigation, reduce delivery accuracy to hundreds of kilometers, and make autonomous navigation a requirement.

\end{abstract}

\section{INTRODUCTION}

\begin{figure}[htp]
    \centering
    \includegraphics[width=9cm]{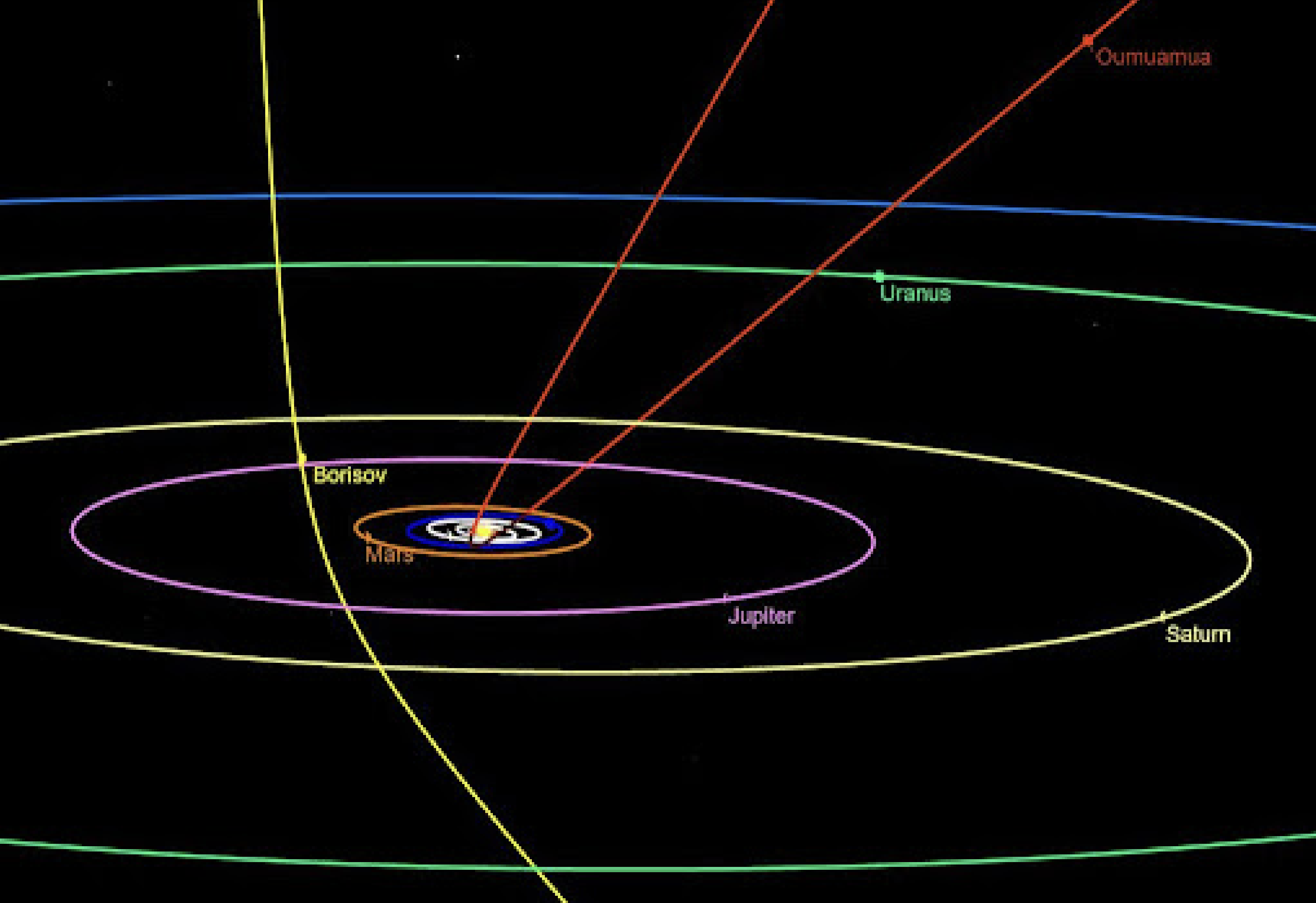}
    \caption{Trajectories of 'Oumuamua and Borisov.}
    \label{fig:isos}
\end{figure}

With the discovery of 1I/`Oumuamua \cite{Meech2017} and 2I/Borisov \cite{borisov}, interstellar objects (ISOs) have been thrust into the forefront of planetary science and the robotic exploration of ISOs is now of significant interest. While there is significant scientific understanding to be gained from observing these objects remotely, ground based observations are far, and in situ characterization remains the most powerful means to studying these objects However, ISOs present unique challenges to mission design and navigation. They by nature pass through our solar system on hyperbolic trajectories as depicted in Figure \ref{fig:isos}. They travel at extreme speeds and spend only a few years inside Saturn's orbit before exiting the solar system forever. Even when the bodies are active and posses comas, they are still dim and thus difficult to detect at distances greater than 10 AU from the Sun, which means there is limited time to respond with an intercept mission. Given these extreme velocity trajectories and limited warning times, rendezvous opportunities with ISOs are likely prohibitive with current and near term propulsion technology \cite{miller2022high}. Thus, the maximum science return would result from a rapid-response fast flyby with possible impactor spacecraft, akin to the Deep Impact mission \cite{ahearn2015} and Comet Interceptor mission \cite{snodgrass2019european}.

However, this rapid-response mission architecture presents its own set of challenges. The hyperbolic orbits drive high relative encounter velocities and high solar phase angles (the angle between vector from target body to Sun and vector from target body to spacecraft), while short observation arc lengths that are intrinsic to a ``rapid-response'' mission produce large ISO position uncertainties that can change dramatically during cruise. The motivation for this work was the concern that high uncertainties during cruise and approach would require prohibitive amounts of delta-v to correct for and limit science return during flyby. To assess this we first establish a population of synthetic ISOs, and then determine which are reachable and the associated spacecraft trajectories. Then we model the uncertainties of all ISOs and spacecraft pairs over the course of each mission starting at launch, and determine the distribution of delta-vs required for navigation and the distribution of final flyby delivery accuracies.

\section{POPULATION MODELING}

\subsection{ISO Population}
In the generation of the ISO population we leveraged previous work by T. Engelhardt et al. \cite{ISOdetection}. From an initial population of 1.7 billion synthetic ISOs, modeled as both active and inert, they derived the ISOs that would have appeared in any Pan-STARRS1 \cite{Wainscoat2016} and Catalina Sky Survey \cite{Christensen2019} telescope fields taken between between 2005-January-1 and 2015-January-1. T. Engelhardt et al. defined as ``detected'' any ISO that resulted in a digest score \cite{Keys2019} greater than or equal to 90. We instead assume that any active ISO is ``detected'' in the first field in which it has an apparent magnitude brighter than 21, i.e., around the current limiting magnitude of the Catalina Sky Survey.
In fact, the detection of cometary activity would ensure that the newly discovered object would get posted on the Possible Comet Confirmation Page\footnote{\url{https://minorplanetcenter.net/iau/NEO/pccp_tabular.html}} and in turn followed up by other observers.
On the other hand, for inert ISOs we additionally require a digest score greater than or equal to 65, which is the threshold that triggers posting on the NEO Confirmation Page\footnote{\url{https://minorplanetcenter.net/iau/NEO/toconfirm_tabular.html}} and, in turn, follow-up.

\begin{figure}[htp]
    \centering
    \includegraphics[width=1.0\linewidth]{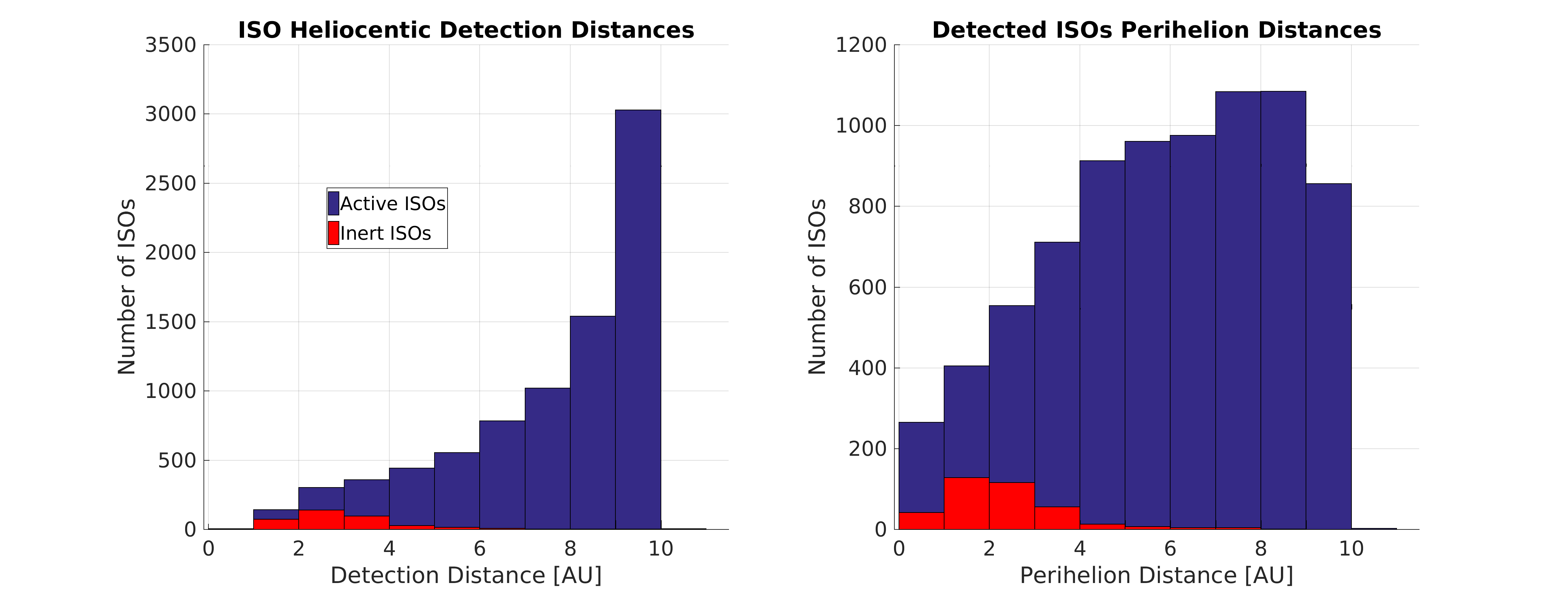}
    \caption{Histograms of detected ISOs' detection distances and perihelion distances.}
    \label{fig:detection distances}
\end{figure}

Of the 1.7 billion synthetic ISOs, 165,378 of them pass within 50 AU of the Sun, and of those 7,813 ISOs pass within 10 AU. Each ISO is studied for detectability as if it were active and as if it were inert. The active ISOs are assumed to become active at 10 AU, becoming significantly brighter and in all cases reaching an apparent magnitude brighter than 21, and thus becoming potentially detectable. When farther than 10 AU from the Sun, with no modeled activity, only a few ISOs reached apparent magnitude 21. All 7813 active ISOs that passed within 10 AU were detected, versus only 371 inert ISOs detected. Figure \ref{fig:detection distances} shows the distribution of distances at which the ISOs are first detected and of the ISOs' perihelion distance. Active ISOs are significantly brighter than inert ISOs and so are detected roughly 2-3 times farther out from the Sun. However, while all the modeled active ISOs maintain detectable magnitudes within 10 AU, the referenced surveys do not cover the entire sky, and so there are trajectory geometries such that a bright ISO does not appear in a field till much later. 2I/Borisov is a good example as it was discovered by an amateur astronomer specifically looking in a region of sky closer to the Sun and thus not covered by current surveys \cite{borisov}.

This methodology produces an ISO population with detection times given current survey telescopes. Future surveys, particularly the Vera C. Rubin Observatory\cite{ivezic2019lsst} and NEO Surveyor\cite{neasurveyor}, will significantly improve our detection capabilities. Vera Rubin will be performing daily full sky surveys down to the 24th magnitude while NEO Surveyor hunts for dim near-Earth objects in the infrared. Future analysis will also have to consider ISO activity and increased brightness outside 10 AU in order to better assess potential detection times at dimmer magnitudes.

\subsection{Spacecraft Trajectory Design}

\begin{figure}[htp]
    \centering
    \includegraphics[width=1.0\linewidth]{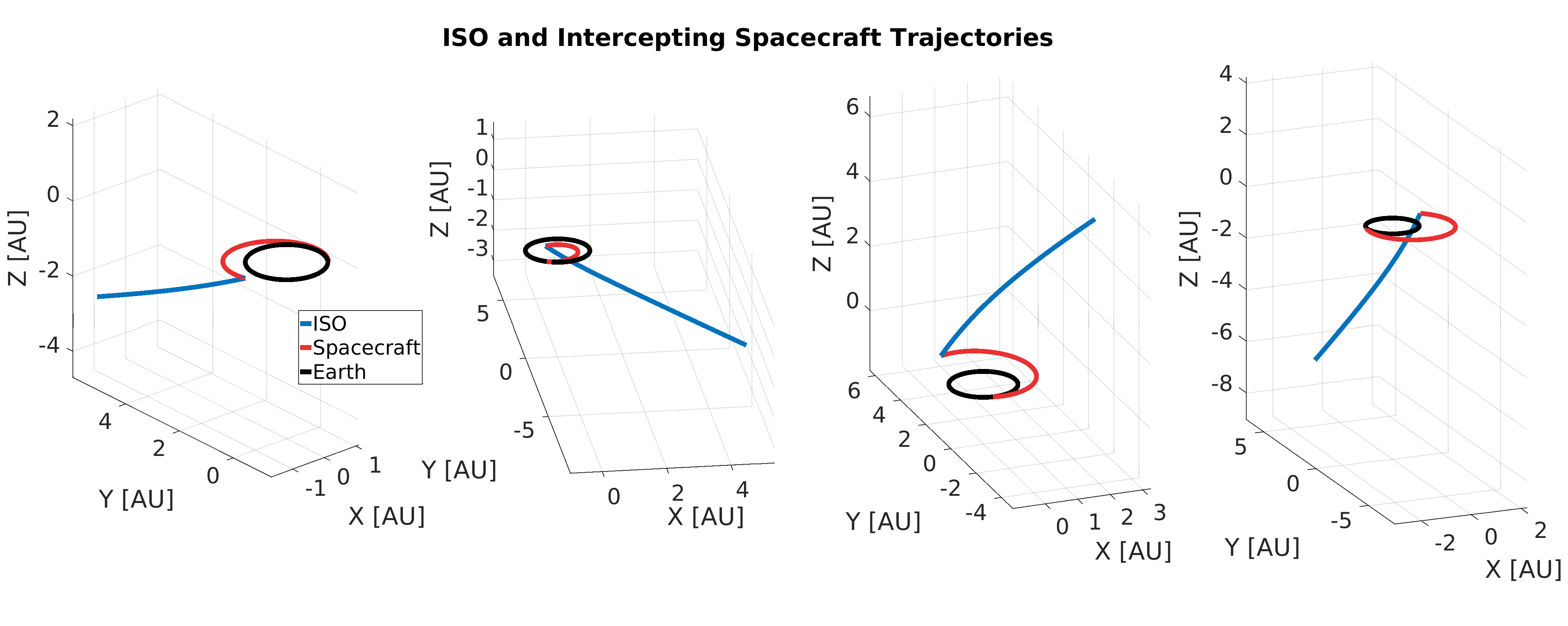}
    \caption{Three example ISO and parent feasible spacecraft trajectories. ISO is plotted starting 1 year out from intercept.}
    \label{fig:example_trajs}
\end{figure}

With the 8,184 total detected ISOs we applied small satellite (SmallSat) system design limitations to constrain a patched-conic search for trajectories that depart Earth to intercept each ISO. The constraints are listed in Table \ref{table:traj_constraints} and outlined in more detail along with the trajectory design process in previous work, ``Interstellar Object Encounter Trade Space Exploration''\cite{IEEE_ISO_trade}. Of the 8,184 detected ISOs, 205 active and 5 inert ISOs can be reached for a fast flyby/impact. While this seems like a negligible fraction, only 931 active and 231 inert ISOs passed within 2.5 AU of the Sun and were thus even within the trajectory design constraints. This gives an intercept chance of 22\% and 2\% respectively for active and inert objects that pass within constraints. Three example ISO and spacecraft trajectories are shown in Figure \ref{fig:example_trajs}. Inclination changes are extremely costly in terms of delta-c, and so the feasible intercept trajectories are largely constrained to the ecliptic. One additional note is that if the launch timing constraint is completely removed, i.e. we no longer care when the ISO is detected, a total of 549 ISOs can be reached for intercept. This is a representative upper bound of possible performance that could be gained from future more powerful observatories.

\begin{figure}[htp] %%%%%% MAKE THIS PLOT DISTINGUISH ACTIVE V INERT
    \centering
    \includegraphics[width=1.0\linewidth]{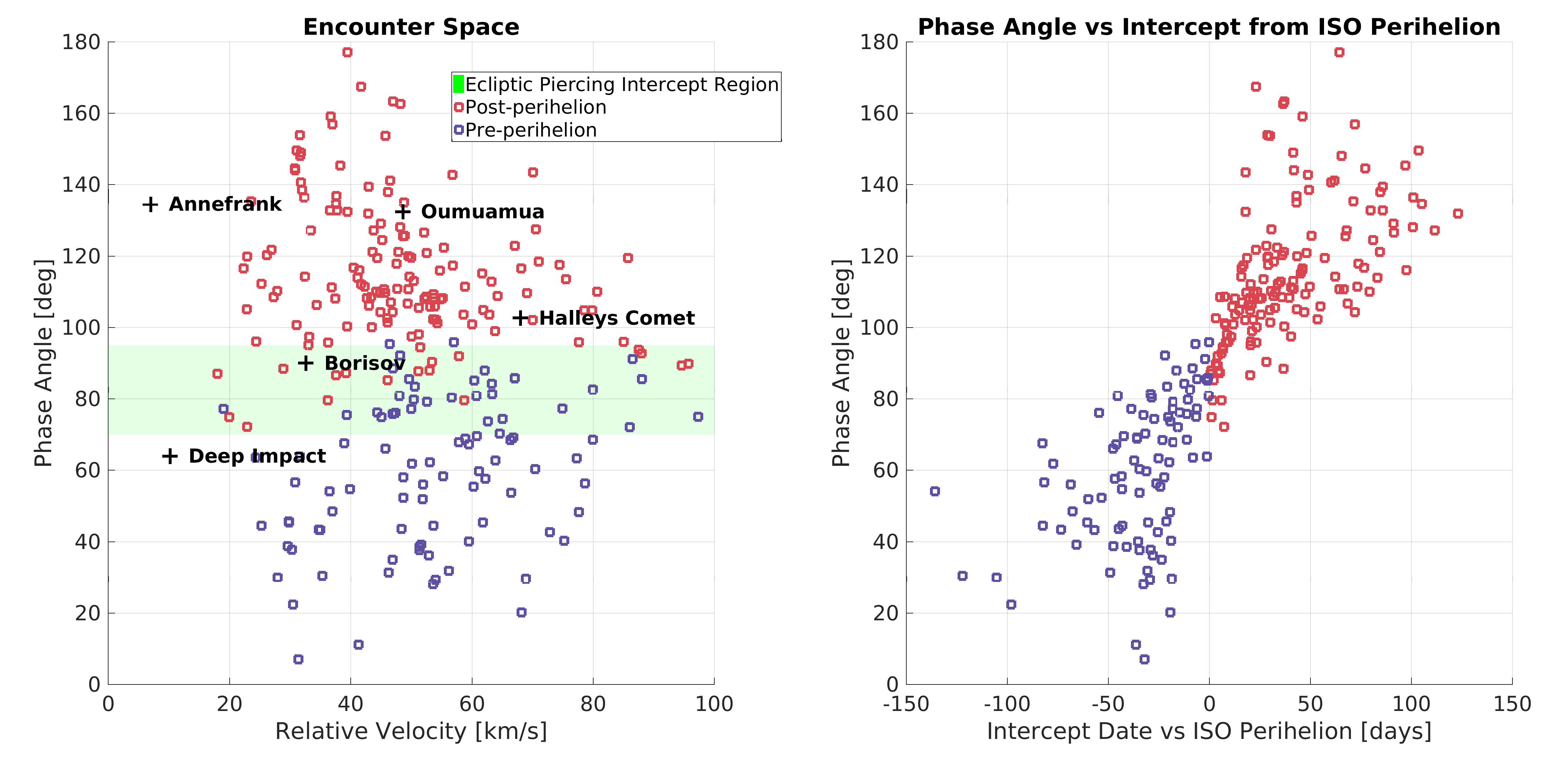}
    \caption{Left, flyby/impact solar phase angle versus relative velocity. Right, flyby/impact solar phase angle versus intercept date relative to ISO perihelion.}
    \label{fig:encouter_space}
\end{figure}

The minimum launch characteristic energy (C3) trajectories produce the encounter space depicted in Figure \ref{fig:encouter_space}. Because of their hyperbolic trajectories, ISOs pass through the inner solar system at extreme heliocentric speeds and at all angles relative to the ecliptic. This means the high speed flyby encounters can be both prograde and retrograde, resulting in the wide range of relative velocities from 20 km/s to nearly 100 km/s. 

With their high speeds, ISOs are traveling significantly faster compared to the flyby spacecraft, given our C3 constraints. As such, the encounter should not be thought of as the spacecraft going to and flying by the ISO. Instead, the spacecraft just seeks to get in the path of the ISO, usually where it pierces or nears the ecliptic. This in turn means that if the ISO is intercepted pre-perihelion, during the inbound portion of its trajectory, the spacecraft will be looking out and away from the Sun, observing a well illuminated target at low solar phase angles. However, if the ISO is intercepted post-perihelion, while outbound, it will typically approach the spacecraft from the direction of the Sun and at high solar phase angles, as highlighted by the right plot in Figure \ref{fig:encouter_space}. Time is a precious resource for rapid response missions, and since post-perihelion intercepts inherently give more time from detection to intercept, they are more likely, representing 61\% of cases, with 50\% of all cases having solar phase angles greater than 100$^{\circ}$. These high solar phase angles will require a highly capable camera and can make it difficult for the spacecraft to image and detect the ISO.

\begin{table}[]
\centering
\caption{Constraints applied to patched-conic trajectory search}
\label{table:traj_constraints}
\begin{tabular}{|l|l|} \hline
Min Time from Detection to Launch   & 10 days                \\ \hline
Max Time from Launch to encounter   & 3 years                \\ \hline
Max orbital revolutions             & 4 before encounter     \\ \hline
Max $C_3$                           & 60 km$^2$/s$^2$          \\ \hline
Max Phase Angle                     & 160$^{\circ}$          \\ \hline
Max Relative Velocity               & 100 km/s             \\ \hline
Max Solar Distance at Encounter     & 2.5 AU                 \\ \hline
Flyby Distance                      & 1000 km                \\ \hline
\end{tabular}
\end{table}

\begin{figure}[htp]
    \centering
    \includegraphics[width=1.0\linewidth]{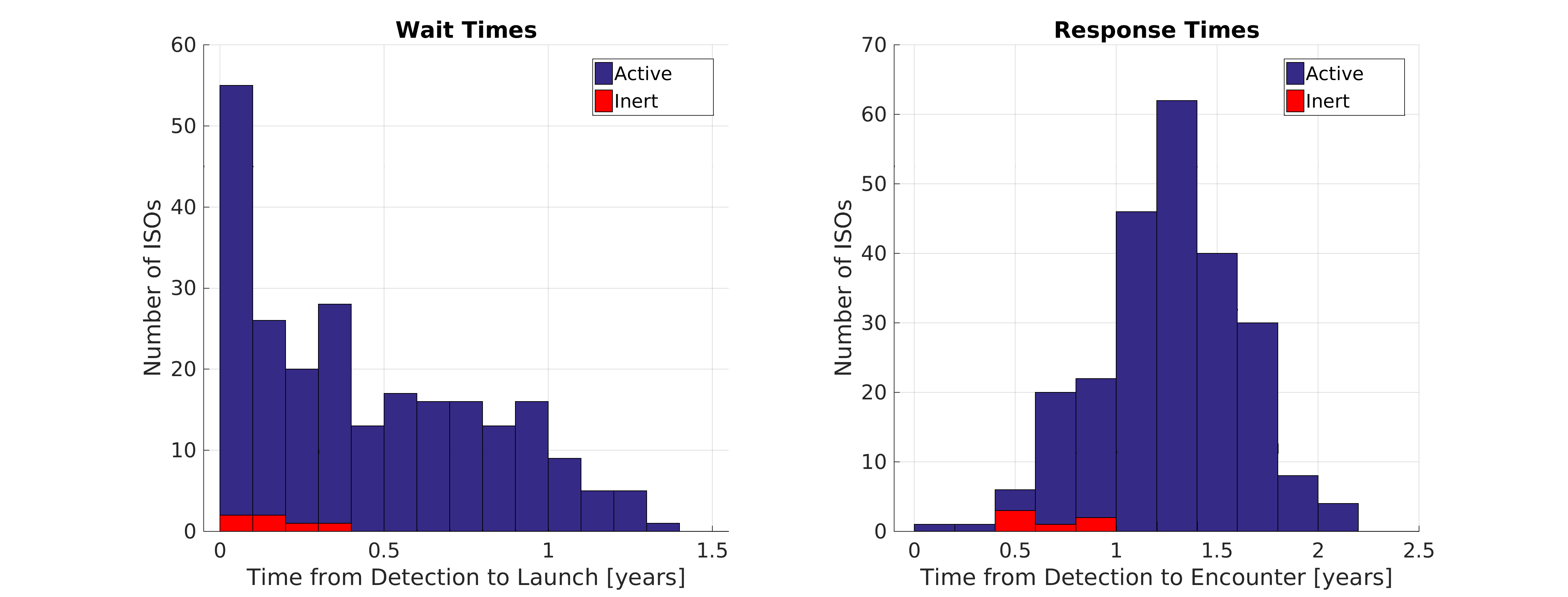}
    \caption{Histogram of spacecraft wait times and response times.}
    \label{fig:sc_times}
\end{figure}

\section{NAVIGATION ANALYSIS}

\subsection{Small Body Deep Space Navigation}

When navigating to a small body in our solar system, the goal is to deliver the spacecraft to a specific flyby location relative to the target. To accomplish this, two key pieces of information need to be estimated: the spacecraft's trajectory and the small body's ephemeris. The spacecraft's orbit determination (OD) is primarily done through tracking data in the form of two-way Doppler, range, and delta-differential one-way ranging with the Deep Space Network (DSN). The target's ephemeris is then computed using primarily optical astrometric measurements from ground-based observatories  and improves as the observation arc length is extended. As the spacecraft approaches the target, optical navigation (opnav), the technique of imaging a target with a camera on-board the spacecraft\cite{owen2008brief}, is also used. Opnav data can directly measure the relative state and improve the target's ephemeris and is powerful due to the fact it is an angular measurement whose metric resolution increases as the spacecraft gets closer to the target. During mission development it is an important step to perform navigation analysis and model the combination of these measurements to predict the uncertainties of the spacecraft and ISO trajectories over time, and make adjustments if needed. The uncertainty over time results are critical to understanding the statistical delta-v required for correction maneuvers and understanding the final delivery accuracies that determine science collection.

The results from ISO population formulation and spacecraft trajectory design ultimately gave 210 unique reachable ISOs and timelines, and for each unique ISO a set of dozens of possible spacecraft trajectories. Typically, navigation analysis involves evaluating a single target body ephemeris with an assumed unchanging ground-based uncertainty, and a single spacecraft trajectory. The challenge this ISO analysis presents is that there are over 200 target ephemerides, each with uncertainties that change dramatically over time -- and also multiple possible spacecraft trajectories. To perform the broad navigation analysis, we first needed to constrain the problem, and so we nominally selected the lowest C3 spacecraft trajectory associated with each ISO. Some alternate trajectories selection methodologies are also discussed in the results. With each spacecraft and ISO pair, the next step was to model their uncertainties over time. The process and the assumptions used is outlined in the following sections.

The last step in our process was to determine the 99th percentile required delta-v (DV99) and final delivery accuracy. This is accomplished by feeding the estimated uncertainty profiles into LAMBIC (Linear Analysis of Maneuvers with Bounds and Inequality Constraints), which produces these statistics by simulating the execution of a sequence of maneuvers through a Monte Carlo process\cite{lambic}. The maneuvers are designed utilizing critical plane targeting, which targets flyby location but not when exactly the flyby will occur. This approach reduces the maneuver design constraints and therefore the required delta-v, especially with high uncertainty targets where relative distance and thus flyby time is poorly known. The coordinate frame depicting delivery results is the b-plane \cite{Kizner1961}; where the $Z$-axis is along the spacecraft-ISO relative velocity vector (here referred to as the downtrack direction), $Y$ axis perpendicular to the relative velocity vector and laying in the ecliptic, and $X$ completing the right-handed frame ($X$ and $Y$ referred to here as crosstrack directions).

\subsection{ISO Uncertainty Modeling}
The ephemeris uncertainty for a discovered ISO affects the feasibility of a mission, especially a rapid response one. We derived the ISO ephemeris uncertainty by simulating optical astrometric observations assuming the cadence described in Table~\ref{tab:cov_inputs}. The astrometric observations were assumed to have an uncertainty of 0.5 arcsec, which is in line with the typical performance of current observing programs \cite{Veres2017}. To mimic a realistic evolution of the ephemeris uncertainty and the real-time uncertainty information that would be available for navigation, we started at the time of discovery and progressively advanced the data cutoff with a time step of one week. For each data cutoff, we computed a best fitting orbit solution and mapped its covariance to compute the future position and velocity uncertainties.

In terms of covariance analysis, one important consideration is the effect of nongravitational perturbations (nongravs), which are typically modeled as radial, transverse, and normal accelerations $A_1 g(r)$, $A_2 g(r)$, and $A_3 g(r)$, respectively, where the $g(r)$ captures the dependency on the heliocentric distance \cite{Marsden1973}. The motion of the first two interstellar objects was significantly affected by nongravitational perturbations, i.e., for `Oumuamua \cite{Micheli2018} $A_1$ was $3 \times 10^{-7}$ AU/d$^2$, while for Borisov \cite{Hui2020} $A_1$ was $7 \times 10^{-8}$ AU/d$^2$. To analyze the dependency of the ISO uncertainty on nongravitational perturbations, we considered three scenarios, similarly to what was done for comet C/2013 A1 (Siding Spring) and the analysis of its close approach to Mars in 2014 \cite{Farnocchia2014}. The first scenario is the optimistic, gravity-only case, i.e., without nongravitational perturbations. As the baseline, best-guess scenario (referred to as ``reference''), we estimate $A_1$, $A_2$, and $A_3$ as part of the orbital fit by assuming a priori uncertainties of $10^{-7}$ AU/d$^2$, $10^{-8}$ AU/d$^2$, and $10^{-8}$ AU/d$^2$, respectively. Finally, as a worst-case scenario (referred to as ``wide''), the $A_1$, $A_2$, and $A_3$ as a priori uncertainties are $10^{-6}$ AU/d$^2$, $10^{-7}$ AU/d$^2$, and $10^{-7}$ AU/d$^2$, respectively. This setup produces three distinct uncertainty evolution profiles for each ISO trajectory. We only consider the gravity-only case for the 5 inert ISOs while the 205 active ISOs are modeled with both reference and wide nongravitational perturbations (nongravs). 

\begin{figure}[htp]
    \centering
    \includegraphics[width=0.49\linewidth]{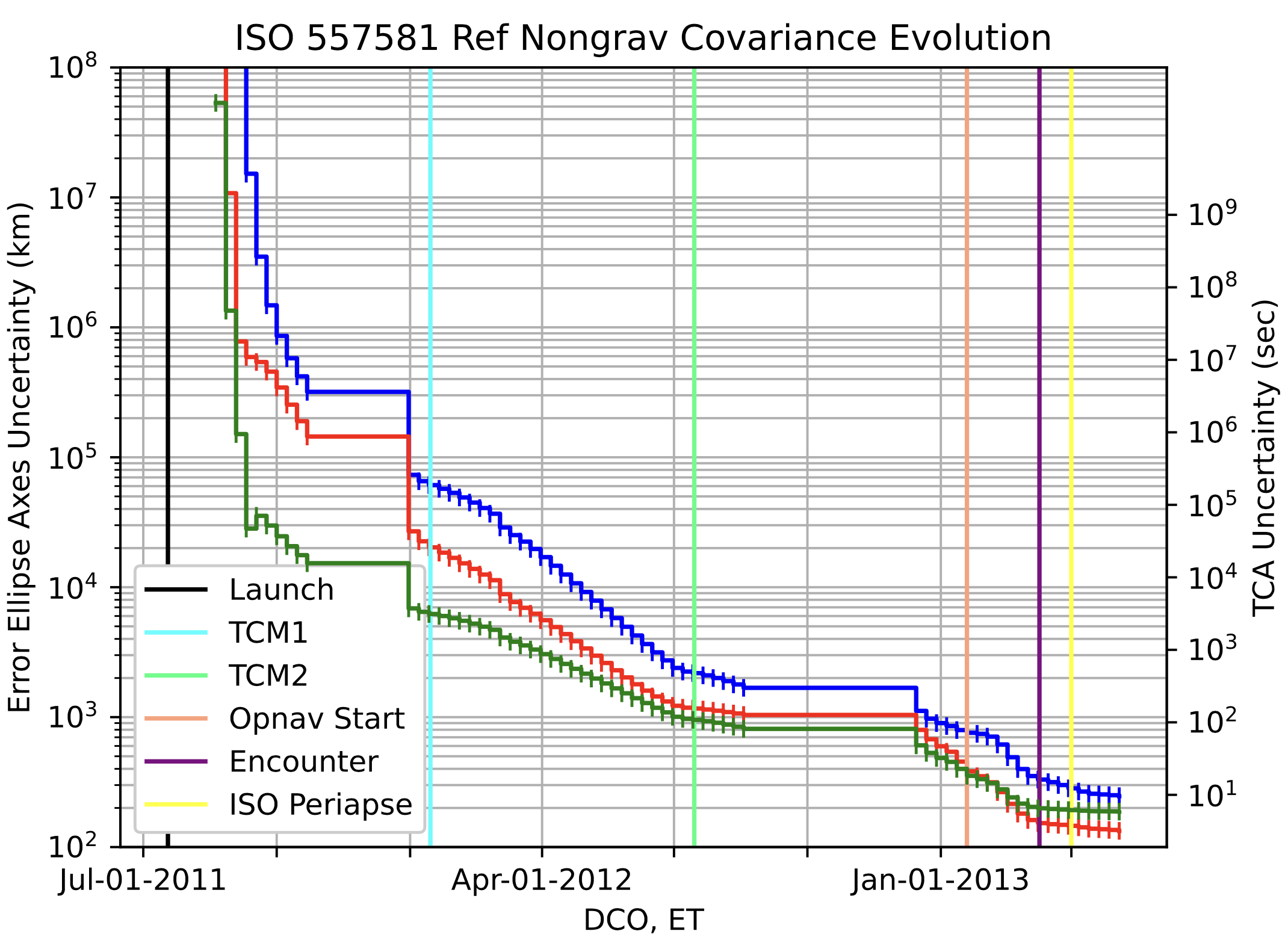}
    \includegraphics[width=0.49\linewidth]{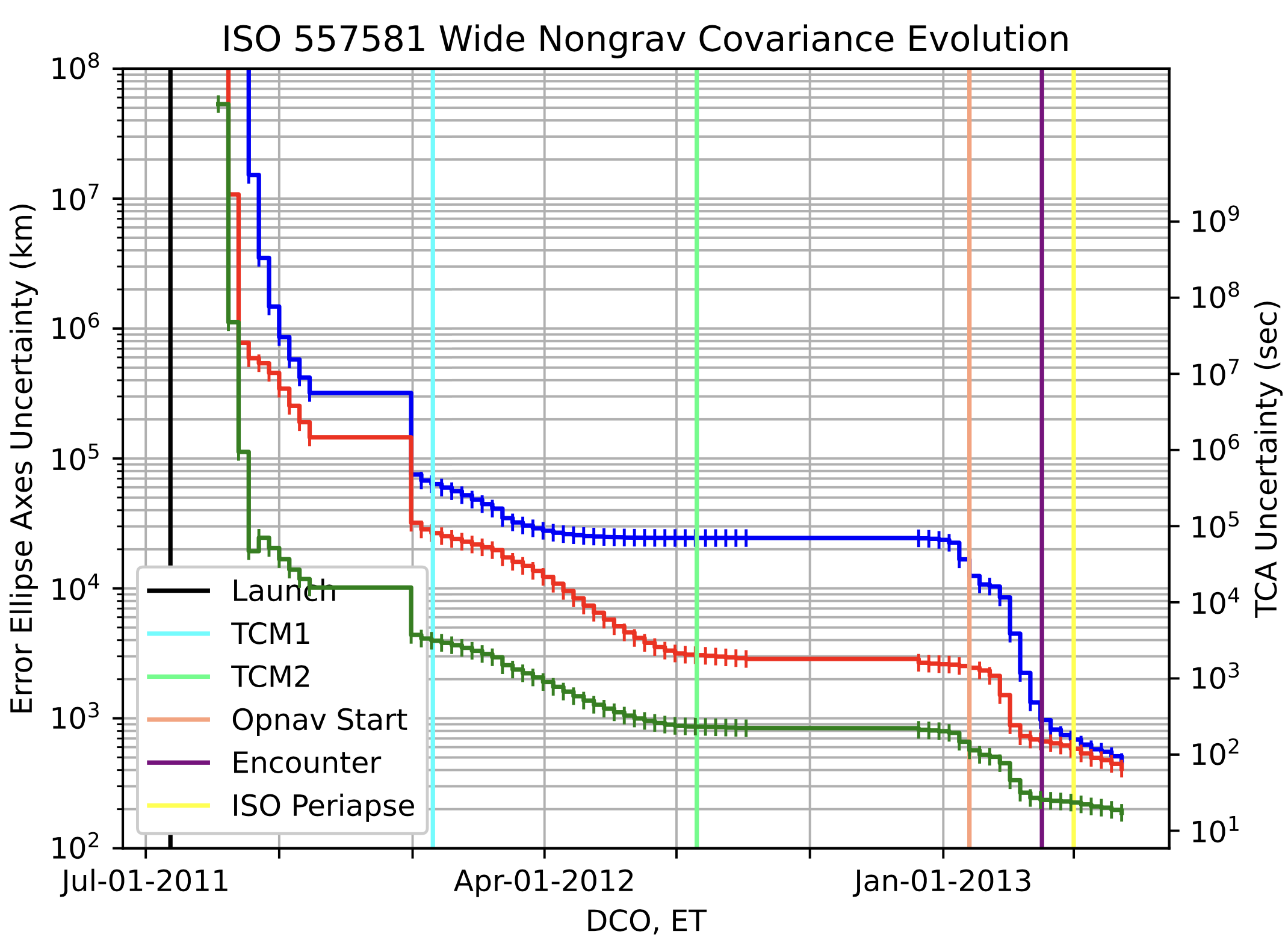}
    \caption{Example ISO 1-sigma ephemeris uncertainty over time when mapped to b-plane at encounter. Left is reference nongravs, right models wide nongravs.}
    \label{fig:cov_evol}
\end{figure}

\begin{figure}[htp]
    \centering
    \includegraphics[width=1.0\linewidth]{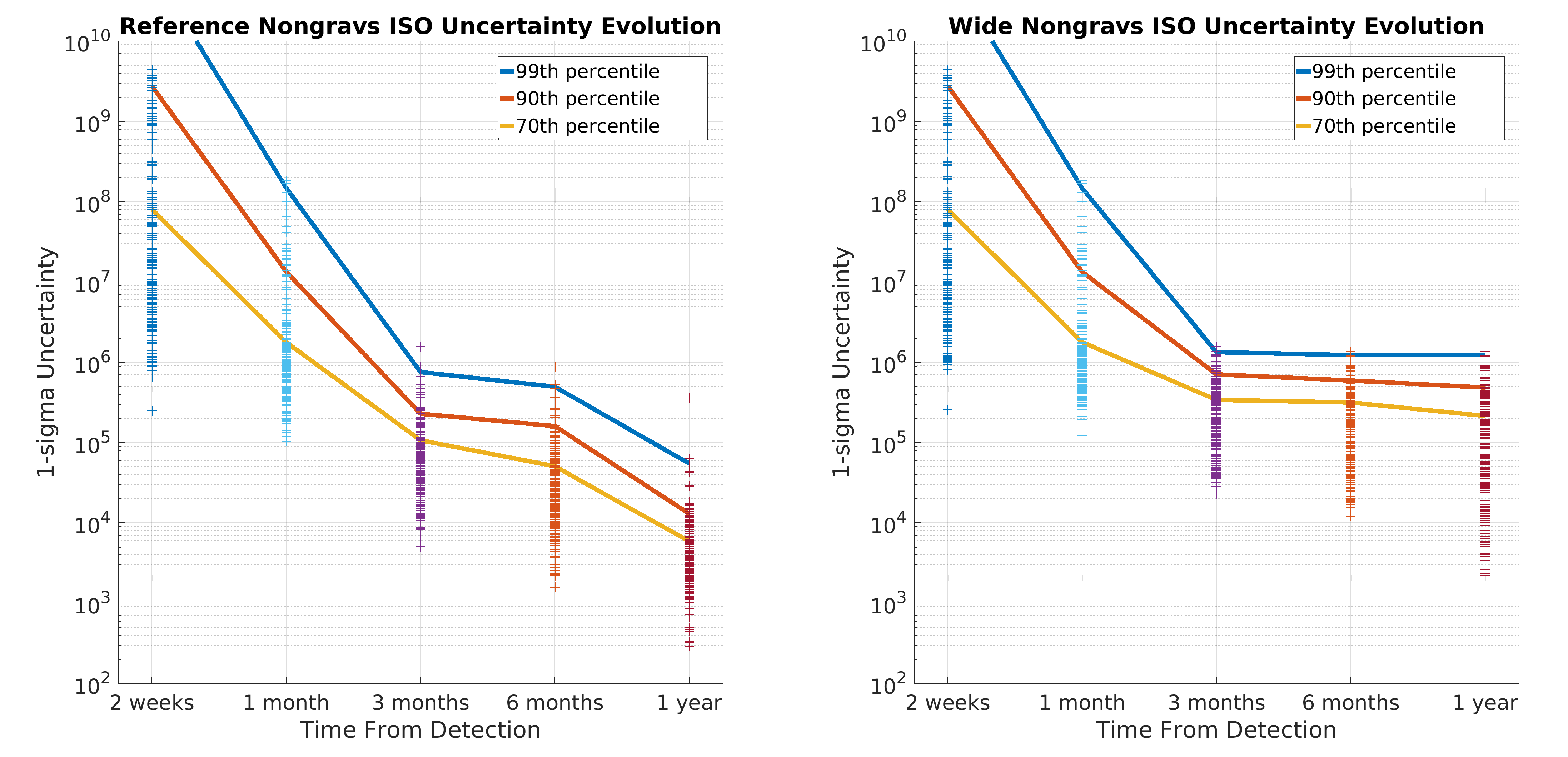}
    \caption{1-sigma uncertainties of where each ISO will be at targeted encounter time versus ISO observation arc-length.}
    \label{fig:iso_uncert_stats.png}
\end{figure}

Figure \ref{fig:cov_evol} shows an example of the resulting ISO uncertainty profiles in the b-plane frame at the spacecraft encounter time versus data-cutoff. This example is indicative of many ISO uncertainty profiles. They start with high initial uncertainties that slowly improve as the observation arc length increases and have unobservable periods than can halt ephemeris improvement. A more statistical representation of the population of ISO uncertainties is then depicted in Figure \ref{fig:iso_uncert_stats.png}. This shows how for both reference and wide nongrav cases, the 90th percentile uncertainty is roughly 10$^8$ km after one month of observation and roughly 10$^6$ km after three months of observation. After that the reference case uncertainties continue to steadily decline -- versus the wide nongrav uncertainties, whose higher magnitude accelerations limit the ability to accurately predict the ISOs location at intercept.

\begin{table}[]
\caption{Inputs to ISO covariance evolution analysis.}
\begin{tabular}{| p{4cm} | p{10.3cm} |}
 \hline
 \multicolumn{2}{|c|}{\textbf{ISO Covariance Setup}} \\ \hline
 Time of Discovery         &  First detection where ISO V magnitude $<$ 21 and solar elongation $>$ 90$^\circ$ \\ \hline
 Observation Schedule   &  Daily observations if V $<$ 24 and elongation $>$ 90$^\circ$, biweekly observations if $24 < V < 26$ and elongation $>$ 60$^\circ$, weekly observations otherwise (as long as $V < 26$ and elongation greater than $60^\circ$)             \\ \hline
 Measurement Uncertainty     &  0.5 arcsec            \\ \hline
 \multirow{2}{7em}{Nongrav a priori Uncertainty}  &  Ref Case: $A_1 = 10^{-7}$ AU/d$^2$, $A_{2,3} = 10^{-8}$ AU/d$^2$ \\ 
 & Wide Case: $A_1 = 10^{-6}$ AU/d$^2$, $A_{2,3} = 10^{-7}$ AU/d$^2$ \\ \hline\hline
 \multicolumn{2}{|c|}{\textbf{Spacecraft Covariance Setup}} \\ \hline
 Initial State Uncertainty  & 100 m, 1 mm/s - position, velocity 1$\sigma$      \\ \hline
 Tracking Schedule          & Four 8 hour Doppler and range passes per week \\ \hline
 OpNav Schedule             & Every 12 hours between E-50 days and E-1 day  \\ \hline
 Measurement Weights        & Two-way Doppler 0.0112 Hz, Two-way SRA range 35 ru, OpNav 1 arcsecond \\ \hline
 Maneuver Schedule          & 1-2 cruise maneuvers timed to minimize $\Delta V$, 3 approach maneuvers at E-40, E-20, E-1 days \\ \hline
 Maneuver Data-cuttof       & 24 hours \\ \hline
\end{tabular}
\label{tab:cov_inputs}
\end{table}

\subsection{Spacecraft Uncertainty Modeling}

\begin{figure}[htp]
    \centering
    \includegraphics[width=13.5cm]{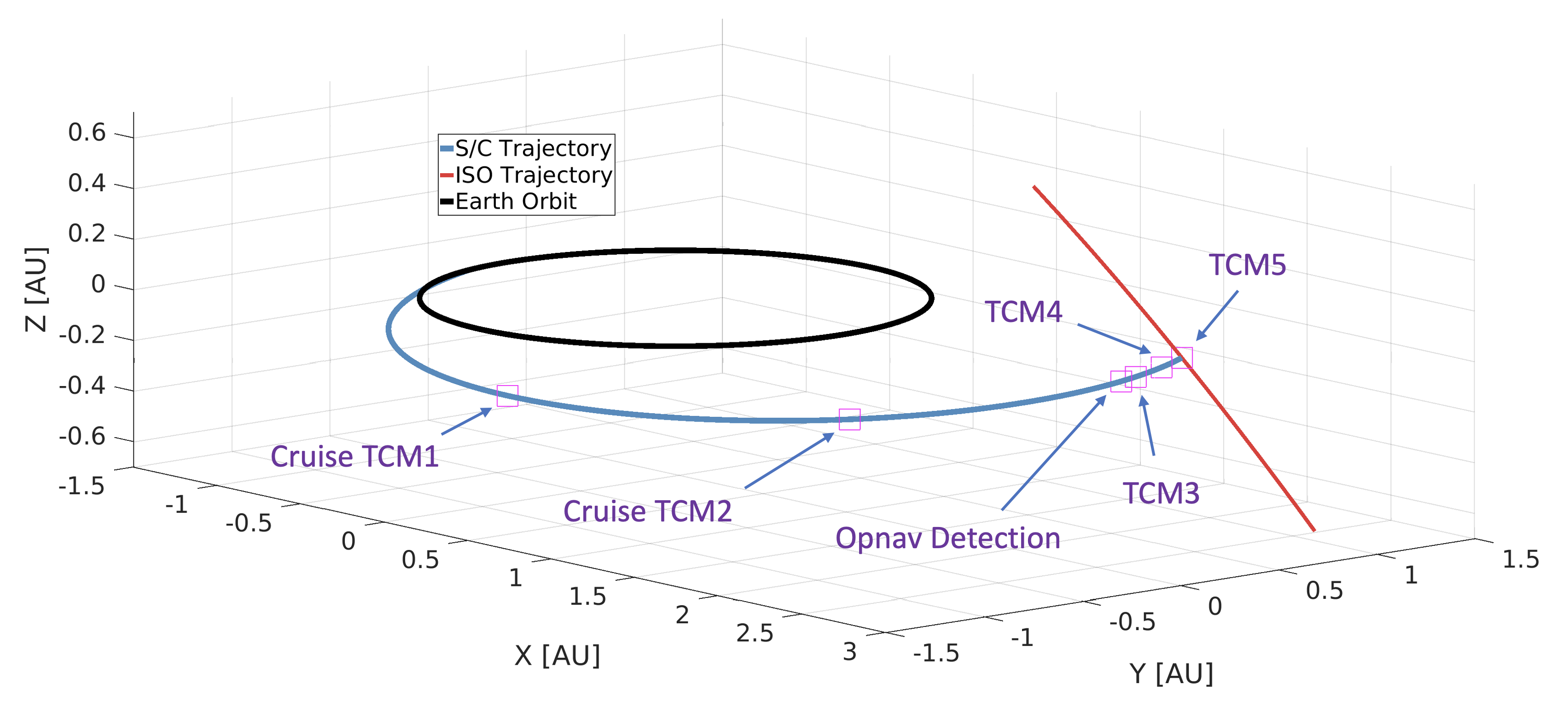}
    \caption{Nominal spacecraft timeline.}
    \label{fig:timeline}
\end{figure}

Each spacecraft's uncertainty is modeled by simulated Doppler, range, and opnav measurements \cite{orbit_determination}. Typically, with a fixed target and trajectory each mission would have its own detailed custom DSN tracking schedule, while the unique geometry and target brightness would determine the opnav imagery schedule. Then any number of trajectory correction maneuvers would be placed throughout the trajectory to maximize performance. However, in order to analyze the 210 spacecraft and ISO pairs, we apply to all a generic data collection and maneuver timing strategy as outlined in Table \ref{tab:cov_inputs} and depicted in Figure \ref{fig:timeline}. The only changes to this schedule were made to investigate the effect of late opnav detection on navigation given most ISO encounters involve high solar phase angles. The emphasis of this analysis is also specifically on how the unique nature of a high speed rapid response mission inflates uncertainties. We therefore chose to only model a simple and benign spacecraft -- estimating only the dynamic state of the spacecraft and ISO while modeling only maneuver execution error and bias solar radiation pressure parameters and no injection error from a launch vehicle. 

For the maneuver timeline, we assumed two trajectory correction maneuvers (TCMs) during cruise -- where cruise is defined as the time from Earth departure to start of opnav. These two maneuvers are initially spaced evenly apart. A check is then done to see if either maneuver is located during a period when the ISO has a low elongation angle and can not be observed. If a maneuver is initially during that period, it is instead placed one week after the next ISO ground astrometry measurement so to be able to take advantage of the improved ephemeris knowledge from the new measurements. This is fairly simple ``optimization'' and possible improvements to maneuver placement is discussed later as future work. During approach there are three maneuvers placed at encounter minus 40 days, 20 days, and 1.5 days, where approach is defined as the time from start of opnav (nominally 50 days from encounter) to the last ground in the loop involvement. 

\section{ACTIVE ISO RESULTS}

\subsection{Delivery Accuracy and Statistical Delta-V}

\begin{figure}[htp]
    \centering
    \includegraphics[width=1.0\linewidth]{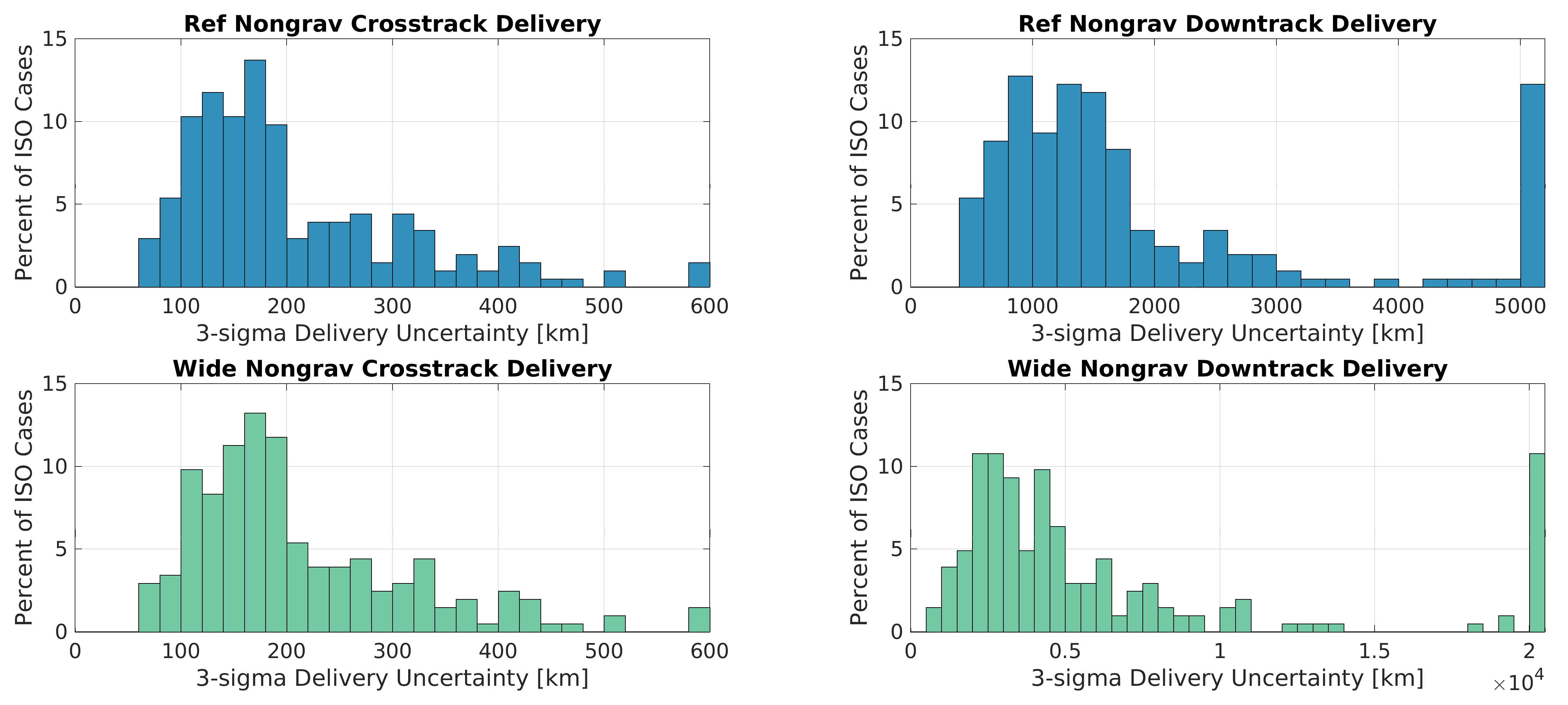}
    \caption{Crosstrack and downtrack delivery uncertainty.}
    \label{fig:active_bplane}
\end{figure}

The resulting delivery accuracies from the final maneuver at 1.5 days from encounter are depicted in Figure \ref{fig:active_bplane} and reveal crosstrack deliveries that are significantly worse compared to typical flyby missions. This is a result of degraded opnav measurement power because the high relative velocities lead to high relative distances. The high crosstrack delivery accuracies are nearly 1:1 correlated with the high relative velocities and this relationship is emphasized by the fact that the larger ISO ephemeris uncertainties from wide nongravs have no effect on crosstrack delivery. The opposite is true regarding the downtrack delivery as opnav cannot measure the distance to a target. The downtrack delivery is therefore largely un-observable and thus defined by the ground-based ISO ephemeris uncertainty. In the reference nongrav case this results in thousands of kilometers of uncertainty, while in the wide nongrav case the uncertainties can be on the order of tens of thousands of kilometers.

These results show that given the assumed optical navigation camera accuracy, the crosstrack delivery error can be a significant percent of the targeted 1000 km flyby distance -- at the extreme cases more than 50\%. The downtrack error results then show that imaging the ISOs during flyby  will likely require autonomous tracking. At closest approach the downtrack uncertainty is fully mapped into the imager or other instrument's FOV -- at 1000 km away this means a downtrack uncertainty of 2000 km represents a roughly 90$^\circ$ long strip of sky. Only with very late autonomous opnav data, when the spacecraft is starting to pass along-side the ISO and the downtrack uncertainty starts to map into the image FOV, can the spacecraft then autonomously estimate when close approach will occur.

\begin{figure}[htp]
    \centering
    \includegraphics[width=11cm]{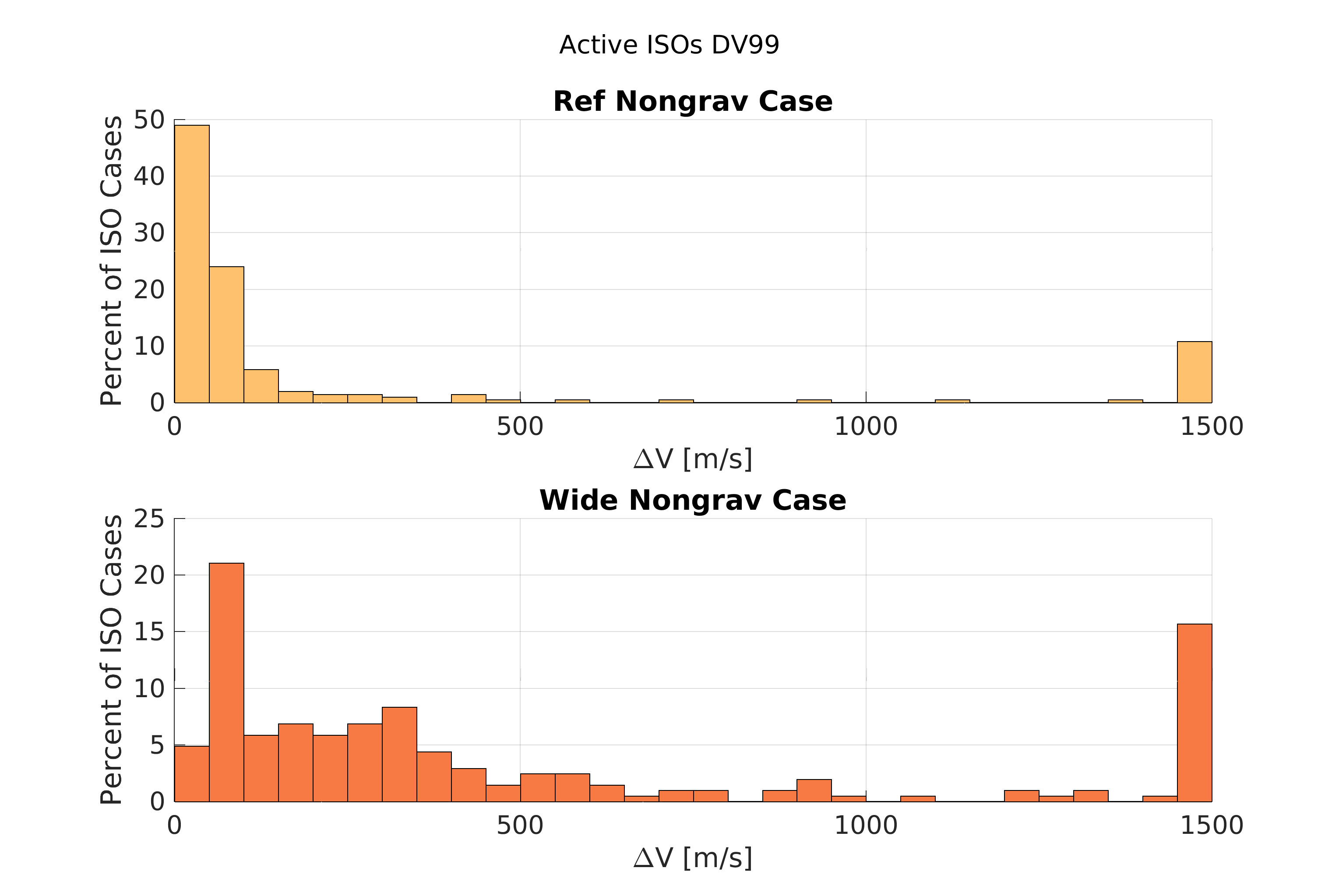}
    \caption{Active ISO DV99s.}
    \label{fig:active_dv}
\end{figure}

A histogram of the required DV99 to navigate to active ISOs is depicted in Figure \ref{fig:active_dv}. These results show that navigating to these highly uncertain targets can require very significant statistical delta-v, particularly in cases with high nongravitational accelerations. With reference nongravs, the majority-75\% of the cases require below 100 m/s of DV99. However, roughly 15\% of cases are shown to require in excess of 500 m/s of DV99. With wide nongravs just 25\% of cases required less than 100 m/s of DV99. The 75th-percentile DV99 case with wide nongravs is shown to be a little over 500 m/s.

\subsection{High Delta-V Cases}

\begin{figure}[htp]
    \centering
    \includegraphics[width=11cm]{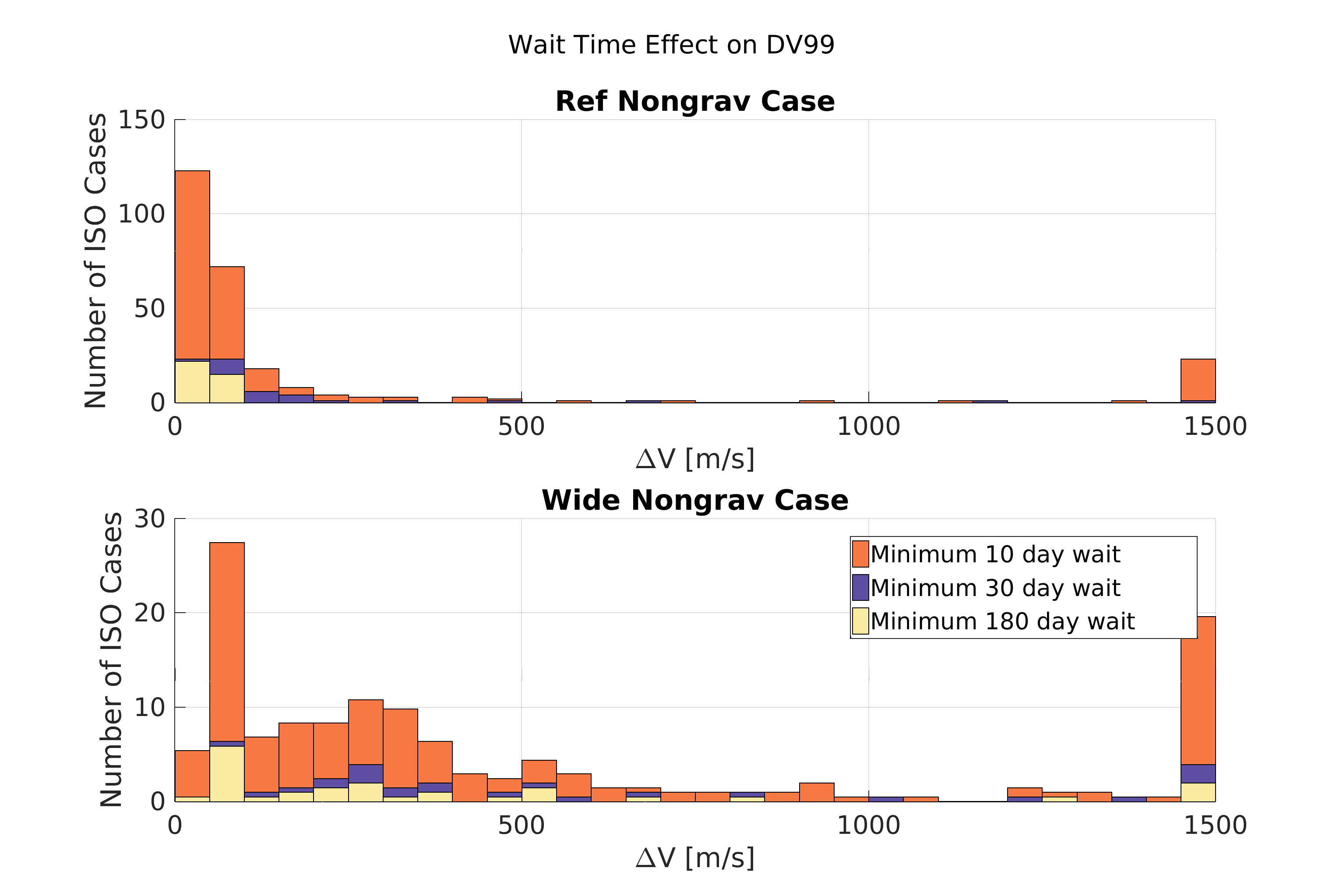}
    \caption{Active ISO DV99s depending for different wait time constraints.}
    \label{fig:wait_dvs}
\end{figure}

While much of the cases' delta-vs fit within a reasonable distribution that tapers off away from the mean, there are a non-negligible $>$10\% number of cases on the edge of the histogram charts that represent multiple kilometer-per-second DV99 cases. These cases require likely infeasible amounts of statistical delta-v. Upon investigation, we determined that the driver of the large DV99 was early launches with a minimum C3 trajectory that departs from Earth only weeks after detection. At this time the ISO's ephemeris is very poorly known as depicted previously in Figure \ref{fig:iso_uncert_stats.png}. Therefore after launch the spacecraft must execute large maneuvers during cruise in order to effectively re-target the new ISO trajectory as more ground data refines its predicted path. Figure \ref{fig:wait_dvs} depicts how requiring longer wait times will remove some of the extreme delta-v cases but will also remove some otherwise viably reachable targets.

By imposing a 30 day wait constraint, 25 spacecraft trajectories no longer minimize C3, and 10 ISOs are no longer reachable. Imposing a 180 day wait results in 37 spacecraft trajectories that no longer minimize C3, and 36\% or 73 ISOs that are no longer reachable. The histogram plots of these new trajectories' DV99s clearly shows almost all of the $>$1 km/s cases have been eliminated and skewed lower. This is completely true with the reference nongrav ISOs, with all 30 day delayed launches below 200 m/s and all 180 day delayed launches below 100 m/s. However, for wide nongrav ISOs the enforced wait times do not remove all of the multi-kilometer-per-second cases, and many still result in DV99s in the hundreds of meters-per-second range. 

\begin{figure}[htp]
    \centering
    \includegraphics[width=0.7\linewidth]{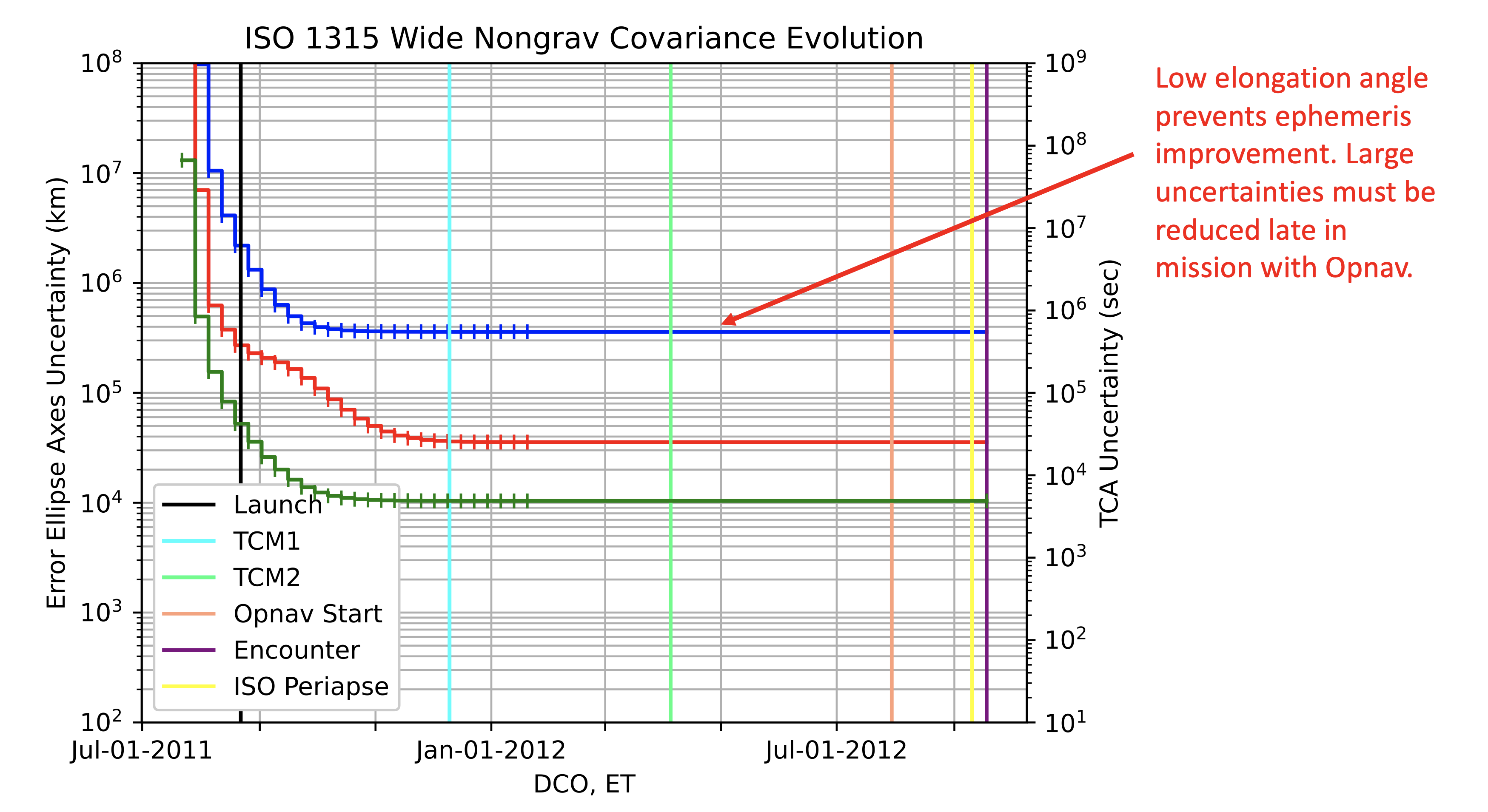}
    \caption{ISO that enters long period of low elongation angle before encounter.}
    \label{fig:high_elong_iso}
\end{figure}

An investigation of those high delta-v, long-wait cases highlighted a secondary reason for large ephemeris uncertainties: ISO low elongation angle observation constraints. Some ISOs, despite being detected far in advance, become un-observable for long periods, when the ephemeris cannot be improved. This happens for a variety of reasons. For example, ISOs approach the inner solar system from a fairly fixed region of sky and so as the Earth's continues on its orbit, the Sun can pass near the ISO's approach asymptote, thus preventing Earth-based observations. Poor timing of a low elongation period can result in high uncertainties at the start of opnav, thus leaving the spacecraft with a significant ephemeris update to make and only 40 days left to make a correction. An example ISO that highlights this behavior is shown in Figure \ref{fig:high_elong_iso}.

\subsection{Late Detection}

Another important navigation challenge to highlight is the possibility of late opnav detection. Late detection means there is less time to correct the ground-based ephemeris, which increase the delta-v costs. For all previous analysis opnav was assumed to start at 50 days from encounter, which requires that the target ISO be bright enough to detect with the onboard imager at that time. As 61\% of encounters involve high solar phase angles this detection may be difficult. For active ISOs, their comas scatter light forward\cite{cometphase} and so should enable early opnav detection even at high phase -- however for low-activity ISOs this may not be the case. We therefore reran the navigation analysis assuming detection 5 days from encounter along with new maneuver assumptions, listed in Table \ref{tab:late_cov_inputs}, to assess how late detection would affect delta-v.

\begin{figure}[htp]
    \centering
    \includegraphics[width=11cm]{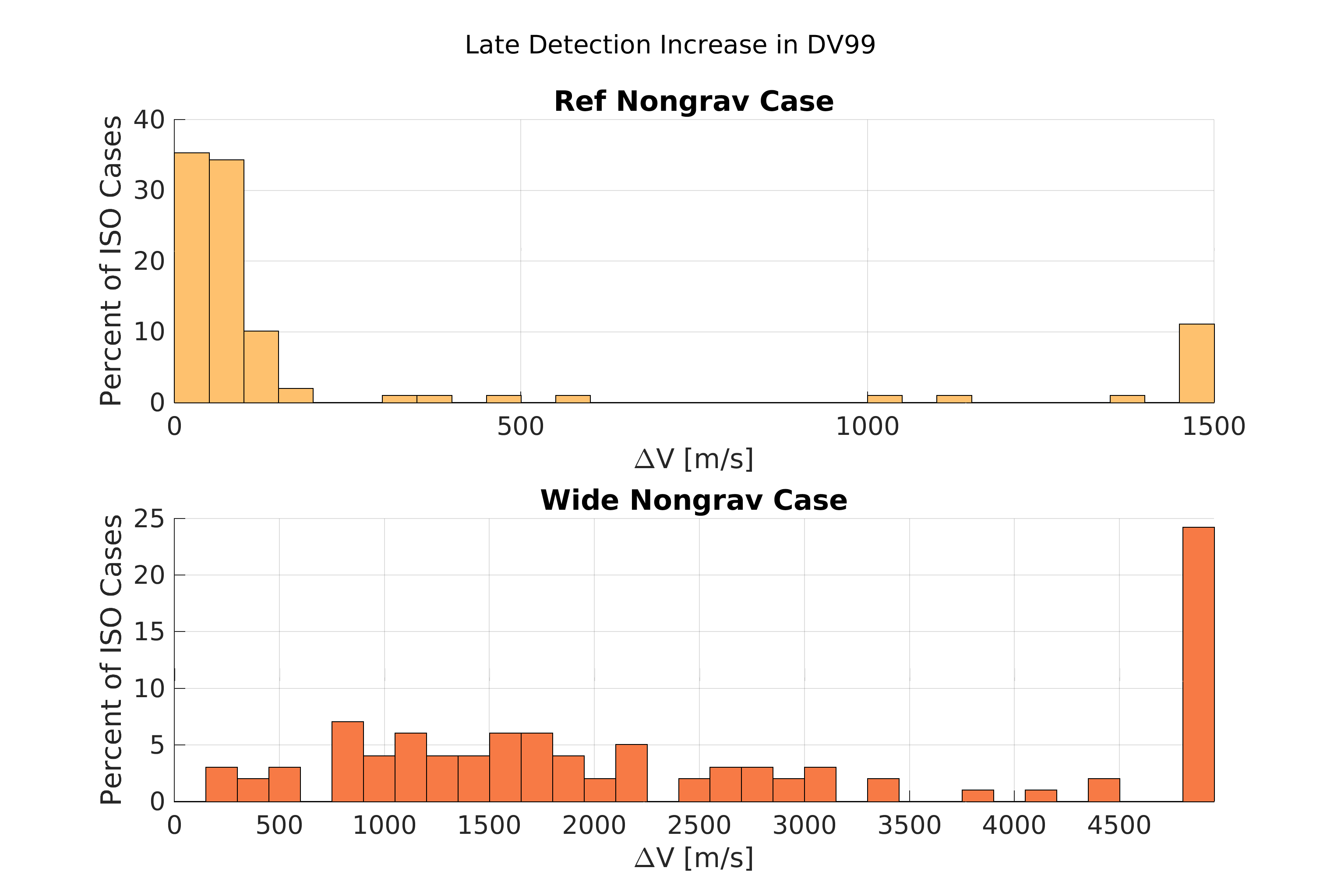}
    \caption{Increase in DV99 due to opnav detection at 5 days versus 50 days.}
    \label{fig:late_detect_dv}
\end{figure}

The 40 day delay in opnav measurement collection results in a modest increase in DV99 for reference nongrav ISOs and an enormous increase in DV99 for wide nongrav ISOs as depicted in Figure \ref{fig:late_detect_dv}. The reference nongrav increase is mostly concentrated around 100 m/s, with some large 1 km/s increases coming from highly uncertain ISOs with substantial gaps in ground observability. For the wide nongrav ISOs, the DV99 increases are almost all above 1 km/s. This highlights how nongravitational forces can severely limit the ISO ephemeris accuracy -- and if the spacecraft is left to correct for that with a maneuver only 4 days from encounter, the delta-v will be massive. Luckily these wide nongrav ISOs are synonymous with highly active ISOs, and so should have detectable coma. 

\begin{table}[t!]
\caption{Late detection new inputs.}
\begin{tabular}{| p{4cm} | p{10.3cm} |}
 \hline
 \multicolumn{2}{|c|}{\textbf{Spacecraft Covariance Setup}} \\ \hline
 OpNav Schedule             & Every 12 hours between E-5 days and E-1 day  \\ \hline
 Maneuver Schedule          & 2 approach maneuvers at E-4 and E-1 days \\ \hline
\end{tabular}
\label{tab:late_cov_inputs}
\end{table}

\section{INERT ISO RESULTS}

\begin{figure}[htp]
    \centering
    \caption{Table of results on 5 reachable inert ISOs.}
    \includegraphics[width=15cm]{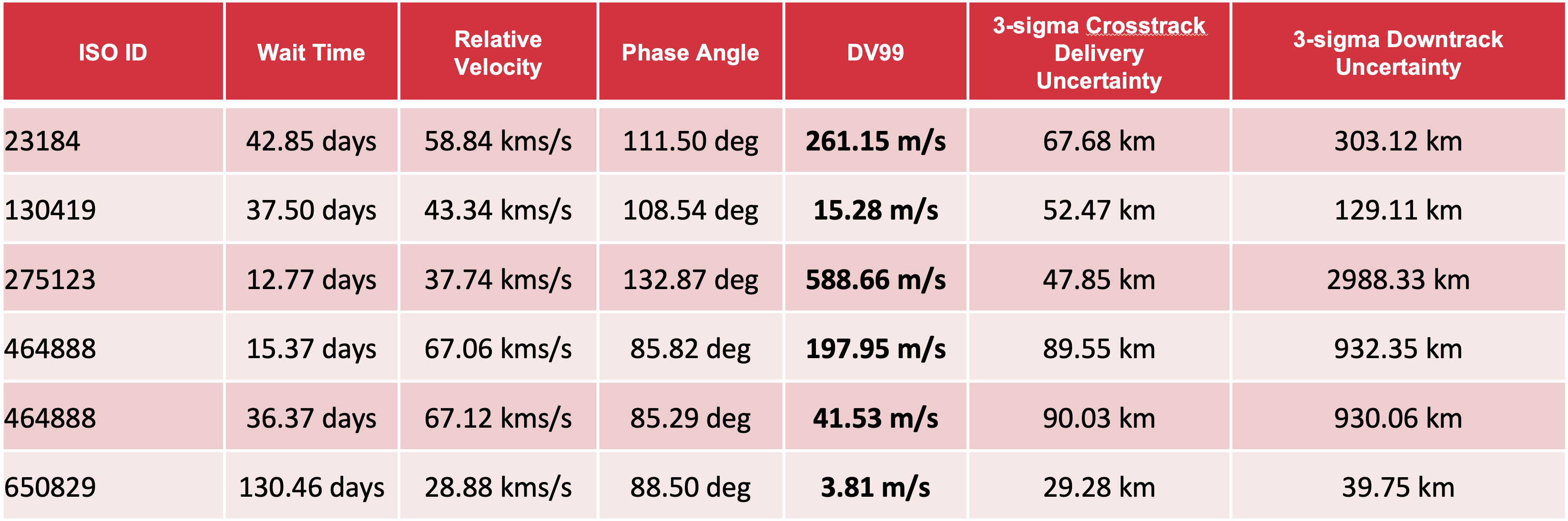}
    \label{tab:inert_isos}
\end{figure}

Even with just five ISOs, each case presents unique mission parameters -- and with the small sample size each of their results can be outlined in more detail. For each inert ISO, with its unique ID, the minimum C3 trajectories with 10, 30, and 180 day wait constraints were analyzed, with results depicted in Table \ref{tab:inert_isos}. The different wait time constraints had no effect except for ISO 464888 -- for which a global minimum C3 trajectory had a wait time of 15 days and still had feasible trajectories after waiting 30 days, unlike for ISO 275123.

ISO 23184 is shown to have a high DV99 even with a wait time of over 30 days. Investigation into the ISO's uncertainty profile showed, as expected, a long gap in ground observation data soon after detection that prevents ephemeris improvement. ISO 130419 is a reasonable target, with a launch wait time over 30 days and good ISO observability and moderate relative velocity -- resulting in a small DV99. ISOs 275123 is the most challenging target with a 13 day wait resulting in a DV99 of 588 m/s. It involves a very high solar phase angle approach which could make opnav imaging extremely difficult. ISO 464888 highlights how waiting 15 days to 30 days can reduce the DV99 from 197 m/s to 41 m/s. And finally ISO 650829 is a nearly idyllic case with a 130 day wait, only 29 km/s relative velocity, and modest solar phase angle.

\section{CONCLUSION}

Our results show that navigating a rapid response mission to an ISO can be extremely challenging. The required delta-v just for statistical trajectory correction maneuvers is comparable to some missions' entire deterministic delta-v. Navigation is rarely considered in early mission development, however for rapid response missions to ISOs these navigation challenges must be modeled from the beginning. If not, the statistical corrections have the potential to break a mission's delta-v budget. There ultimately may be ISOs that, while accessible from a trajectory design perspective, may not be feasible targets due to extreme navigation delta-v.

The high relative velocities are also shown to limit delivery accuracies to hundreds of kilometers. To significantly improve this, autonomous OD and maneuver execution is likely needed \cite{mages2022navigation}. High ISO ephemeris uncertainties are also shown to massively inflate the downtrack uncertainties during flyby. Meaningful science collection will therefore likely require autonomous tracking capabilities.

Other conclusions from this analysis include the requirement for a dynamic navigation team that is closely integrated with the ground-based ISO OD team. Nongravitational accelerations are also shown to be possibly the most critical parameter in determining navigability. A highly capable opnav imager is also needed for long range detection even at high solar phase angles. And finally, a counter-intuitive result is that for a rapid response mission, rather than launching to the target as soon as possible, it can instead be more optimal to launch as late as possible.

\section{ACKNOWLEDGEMENTS}

The research described in this paper was carried out at the Jet Propulsion Laboratory, California Institute of Technology, under a contract with the National Aeronautics and Space Administration. Copyright (c) 2022 California Institute of Technology. U.S Government sponsorship acknowledged.

\bibliographystyle{AAS_publication}   % Number the references.
\bibliography{references}   % Use references.bib to resolve the labels.

\end{document}